\documentclass[prx,onecolumn,showpacs,amsmath,amssymb,superscriptaddress]{revtex4-2}
\usepackage{graphicx}% Include figure files
\usepackage{dcolumn}% Align table columns on decimal point
\usepackage{bm}% bold math
\usepackage{hyperref}% add hypertext capabilities
\hypersetup{colorlinks,citecolor=blue, filecolor=blue, linkcolor=blue , urlcolor=blue}
\usepackage{ulem}
\usepackage[version=3]{mhchem}
\usepackage{hyperref}

\begin{document}

%\title{Spin-splitting of excitons in WSe$_2$/CrI$_3$ Van der Waals heterostructures}
\title{Challenges and opportunities in proximity-driven %IZ - added
exciton-spin engineering in van der Waals heterostructures}
%\title{Exploring the controllability of exciton-spin engineering in WSe$_2$/CrI$_3$ van der Waals heterostructures}

\author{Mushir Thodika}    
%\email{mushir.thodika@nrel.gov}
    \affiliation{National Renewable Energy Laboratory, Golden, CO, 80401 USA}  
\author{Dimitar Pashov}
\affiliation{King’s College London, Theory and Simulation of Condensed Matter, The Strand, WC2R 2LS London, UK}
\author{Igor \v Zuti\' c}
    \affiliation{Department of Physics, University at Buffalo, State University of New York, Buffalo, NY 14260 USA }
\author{Mark van Schilfgaarde}
    \affiliation{National Renewable Energy Laboratory, Golden, CO, 80401 USA}
\author{Swagata Acharya}
%\email{swagata.acharya@nrel.gov}
    \affiliation{National Renewable Energy Laboratory, Golden, CO, 80401 USA} 

\begin{abstract}
\begin{center}
  \textbf{Abstract}\\  
\end{center}

van der Waals heterostructures consisting of transition metal dichalcogenides (TMDs) and two-dimensional (2D) magnets offer a versatile platform to study the coexistence and transformation of different excitons. By focusing on TMD WSe$_2$ and 2D magnetic CrI$_3$, as a bilayer WSe$_2$/CrI$_3$ and a trilayer CrI$_3$/WSe$_2$/CrI$_3$, we provide their description using a parameter-free, high-fidelity many-body perturbation theory. This {\em ab initio} approach allows us to elucidate the character of magnetic Frenkel excitons in CrI$_3$ and how the nonmagnetic Wannier-Mott excitons in WSe$_2$ are modified by the proximity of CrI$_3$. We reveal novel proximity-induced interlayer excitons in these heterostructures. In contrast to the sensitivity of proximity-induced modifications of excitons in WSe$_2$, which depend on the interfacial details, the interlayer magnetic excitons are remarkably \textcolor{black}{robust and are present across the different stacking configurations between WSe$_2$ and CrI$_3$}, simplifying their experimental demonstration. These findings suggest unexplored opportunities for information transduction using magnetic excitons and integrating photonics, electronics, and spintronics in proximitized materials. 

\end{abstract}
\maketitle
%%%%%%%%%%%%%%%%%%%%%%%%%%%%%%%%%%%%%%%%%%%%%%%%%%%%%%%%%%%%%%%%%%%%%
%% Start the main part of the manuscript here.
%%%%%%%%%%%%%%%%%%%%%%%%%%%%%%%%%%%%%%%%%%%%%%%%%%%%%%%%%%%%%%%%%%%%%
\section{Introduction}
Modern day quantum photonics, optoelectronics, neuromorphic computing, spintronics, %IZ
and straintronics rely heavily on functional heterostructure materials that can act as building blocks for the integrated circuitry~\cite{pham20222d,kroemer2005heterostructure,Zutic2004:RMP}. A functional heterostructure for such applications 
requires %IZ s added
two crucial properties: a) it should be small and integrable and b)  its functional properties can be controlled in-situ. 
The recently discovered two-dimensional (2D) %IZ - added 2D introduced
van der Waals (vdW) %IZ added
materials~\cite{liu2016van,geim2013van,chhowalla2013chemistry,huang2017layer,seyler2018ligand} have revolutionized the heterostructure functionality as they can be scaled down from bulk to a few nanometers and a few layers. At the nanoscale, in-situ tuning of heterostructure properties becomes possible, and also the remarkable principle of proximity-induced quantum phenomena emerges~\cite{zuticpe1,montero2002nanostructures,deutscher2018proximity}.  Together, they promise a new era of efficient nanoscale quantum integrated circuitry~\cite{fiori2014electronics,jariwala2014emerging,zhang2015ultrathin}. %IZ ~ added
The functionality of semiconductor industry has historically relied on doping~\cite{shklovskii2013electronic,schubert2015doping,zakutayev2014defect,furdyna1988diluted}, alloying~\cite{ning2017bandgap}, defect formation~\cite{PhysRev.164.1043,doherty2013nitrogen,PRXQuantum.3.010339,bulancea2021dipolar,gali2019ab,PhysRevB.106.014202,PhysRevMaterials.5.084603}, %IZ formation  Moved before
and chemical reactions~\cite{voiry2015covalent,yan2012chemistry}. These approaches have been used to tailor semiconductor properties over a large range of the electromagnetic spectrum, typically achieving desired properties, logic, memory, sensing, and coherent emission, from ultraviolet to infrared and telecommunication window. However, in-situ control of these properties through doping or alloying is a daunting task and has faced significant challenges. Exploiting the proximity effects %IZ s added, there could be multiple proximity effects in a single heterostruct.
in a vdW %IZ van der Waals 
heterostructure provides a fresh alternative %IZ text added  
where, unlike during the decades of prior work~\cite{Hauser1969:PR},
it can be crucial to provide the description of proximity effects  beyond the single-particle picture~\cite{Scharf2017:PRL}.
%{van} %IZ
%der Waals 
vdW %IZ maybe use the abbreviation?
materials host a spectrum of excitonic transitions which are often well split off from the band edges and are not contaminated by the continuum of electron- and hole-like excitations. Controlled tuning of the excitonic properties holds the key to their functionality. In transition metal dichalcogenides (TMDs) %IZ s
and hexagonal boron nitride (hBN) a series of such excitons can be realized over the visible and ultraviolet window. Several of these excitons have large binding energies ranging up to a few 100's of meV. However, they are nonmagnetic and, hence, the main body of work focuses on tuning the valley~\cite{norden2019giant,ge2022enhanced,seyler2018valley,zhao2024magnetic,heissenbuttel2021valley} or pseudo-spin~\cite{hu2020manipulation,zollner2019proximity} degrees of freedom through the proximity effect. It is only recently that an independent field of research has emerged where these %IZ TMD's 
TMDs
are functionalized with magnetic vdW %IZ van der Waals 
materials to access the microwave and infrared resonant and non-resonant energy windows. vdW %IZ Van der Waals
magnets with large frozen spin moments can generate a proximity field in-situ and split the otherwise nonmagnetic excitonic states of the TMD. The functionality of the heterostructure relies on the ability to induce magnetic field and split the excitons by desired amounts. 
%Recent works have focused primarily on the absolute magnitudes of the spin-splitting.  

There are several outstanding problems, nevertheless, that complicate this process. Both TMDs %IZ TMD's 
and the vdW magnets have rich excitonic spectra~\cite{pollini1970intrinsic,grant1968optical,dillon1966magneto,zhang2019direct,ye2014probing,qiu2013optical} and the two often reside at comparable energies. While the TMD nonmagnetic excitons are extended bosonic states with bohr radii of several dozen nanometers~\cite{PhysRevB.93.235435}, the magnetic excitons in vdW magnets can be localized within one unit cell~\cite{acharya2022real,wu2019physical} (bohr radius of %IZ
half a nanometer).  Wavefunctions of the Frenkel excitons in the vdW magnets are primarily determined by the onsite dipole forbidden $dd$ transitions while the Wannier-Mott excitons in TMDs %IZ TMD's 
are mostly dipolar $pd$ in character.  Proximity-generated interactions between these two extremes induce intricate interplay between them, affecting their coherence, oscillator strengths,  and the %IZ
extent of the wavefunctions.  Equally intricate is the question of how a proximity-generated %IZ - added
magnetic field from the vdW magnet splits and spreads over the extended TMD excitonic state.  
%All of these effects have yet to be understood.
%\begin{figure}[hbt!]
%    \centering
%    \includegraphics[width=\textwidth]{Figures/summaryfig.png}
%    \caption{An illustration depicting the origin of different excitonic features in a H-WSe$_2$/CrI$_3$ heterostructure. Cr \textit{d} states energetically split into a degenerate set of three $t_{2g}$ states and two $e_g$ states in the presence of an octahedral crystal field. The relative alignment of CrI$_3$ atomic-like states and the WSe$_2$ bloch states as predicted by $\mathrm{QS}G\hat{W}$ ensures that the primary excitonic features from WSe$_2$ are preserved in both spin channels. At the same time, the important excitonic features from the CrI$_3$ layer are also preserved in the heterostructure. In addition to the excitons originating from the respective layers, a new set of magnetic interlayer %IZ excitons are observed with enhanced oscillator strengths compared to the relatively dark CrI$_3$ excitons and the energies of these excitons are highly robust in both the bilayer and trilayer configurations.}
    %\label{fig:summary}
%\end{figure}
A sound basis for insights into these intricate phenomena requires an \textit{ab initio} approach.  
%Time dependent density functional theory (TDDFT) is the simplest that incorporates excitonic effects, but its fidelity is doubtful in current implementations. 
Diagrammatic many-body perturbation theory (MBPT) has achieved great success in describing excitons in TMDs since non-local charge correlations within such a framework are satisfactorily described~\cite{PhysRevB.93.235435,PhysRevLett.111.216805}.  It retains the momentum dependence of Coulomb %IZ
%- maybe not needed (just as - was not used later)
correlations, which is sufficient for Wannier-like electron-hole attraction, as they are well characterized in terms of Bloch states.  MBPT %IZ spell out
is also partially successful in characterizing excitons in the Frenkel (atomic) limit, e.g. deep excitons in vdW magnets~\cite{acharya2022real,shao2024exciton,datta2024magnon}.  Here excitons are largely atom-local, so the dynamic (frequency dependent) structure of the Coulomb correlation is more important than its momentum dependence.  Partially filled atomic $d$ states suffer from stringent orbital- and spin-angular momentum selection rules (Laporte rule)~\cite{laporte1925some} that can prohibit optical transitions.  Frenkel excitons are nevertheless detected in both absorption and photoluminescence emission spectroscopy, primarily owing to the fact that the optical selection rules of a single atom are relaxed in presence of the ligands that lead to $dp$ mixing and the symmetry lowering mechanisms mediated primarily by the phonons~\cite{acharya2022real}.  These excitons can be captured in modern \textit{ab initio} MBPT, but only partially.  In particular, a spin-flip onsite $dd$ transition  involves diagrams not included in current implementations of MBPT, and it remains one of the most complicated objects to characterize.  Ligand-field approaches and the resulting Tanabe-Sugano ($TS$) diagrams\cite{tanabe1954absorption,tanabe1956absorption} do provide a guideline for the atomic multiplet transitions, both spin flip and non spin-flip $dd$ transitions, which can be both excitonic and bi-excitonic in nature. However, these $TS$ diagrams are phenomenological models and can not be used for predictive purposes.

In recent times, we have built a higher order extension of quasiparticle self-consistent \textit{GW} approximation~\cite{qsgw,questaal_paper}, called $\mathrm{QS}G\hat{W}$~\cite{Cunningham2023}, where the random phase approximation (RPA) is augmented with a vertex made of ladder diagrams, which improve the screening and incorporates excitonic effects. QS$G\hat{W}$ is an extension of QS$GW$ where electronic eigenfunctions are computed self-consistently, including excitonic correlations in $\hat{W}$. Self-consistency is imposed for both the self-energy $\Sigma$ and the charge density. The latter is usually neglected in \textit{GW}, however, it has been demonstrated to modify the electronic structure for a certain class of material systems  \cite{acharya2021importance,tise2}. Self-consistency in $\Sigma$ ensures that feedback between the one-particle spectrum and the Coulomb interactions is accounted for~\cite{Vidal10}.  This is particularly important when magnetic degrees of freedom are involved, as there is an additional coupling between spin and $\Sigma$.  $G$, $\Sigma$, and $\hat{W}$ are updated iteratively until all of them converge.  Our results are thus parameter-free and have no starting point bias. The theory and its application to a large number of both weakly and strongly correlated insulators is given in Ref.~\cite{Cunningham2023}. This work and subsequent ones have established that QS$G\hat{W}$ is uniformly high fidelity in both the 1-particle and 2-particle sectors.
%, provided spin fluctuations or the electron-nuclear interactions are not significant. 
As an additional step, QS\textit{GW} can be augmented with dynamical mean-field %IZ - added
theory (DMFT)~\cite{georges1996dynamical,kotliar2006electronic} where all local vertices for spin and charge fluctuations are built in.  This allows for spin-flip atomic multiplets to be captured, not incorporated in low-order MBPT.~\cite{acharyaTheoryColorsStrongly2023}.

\begin{figure}[htb!]
    \centering
    \includegraphics[width=\textwidth]{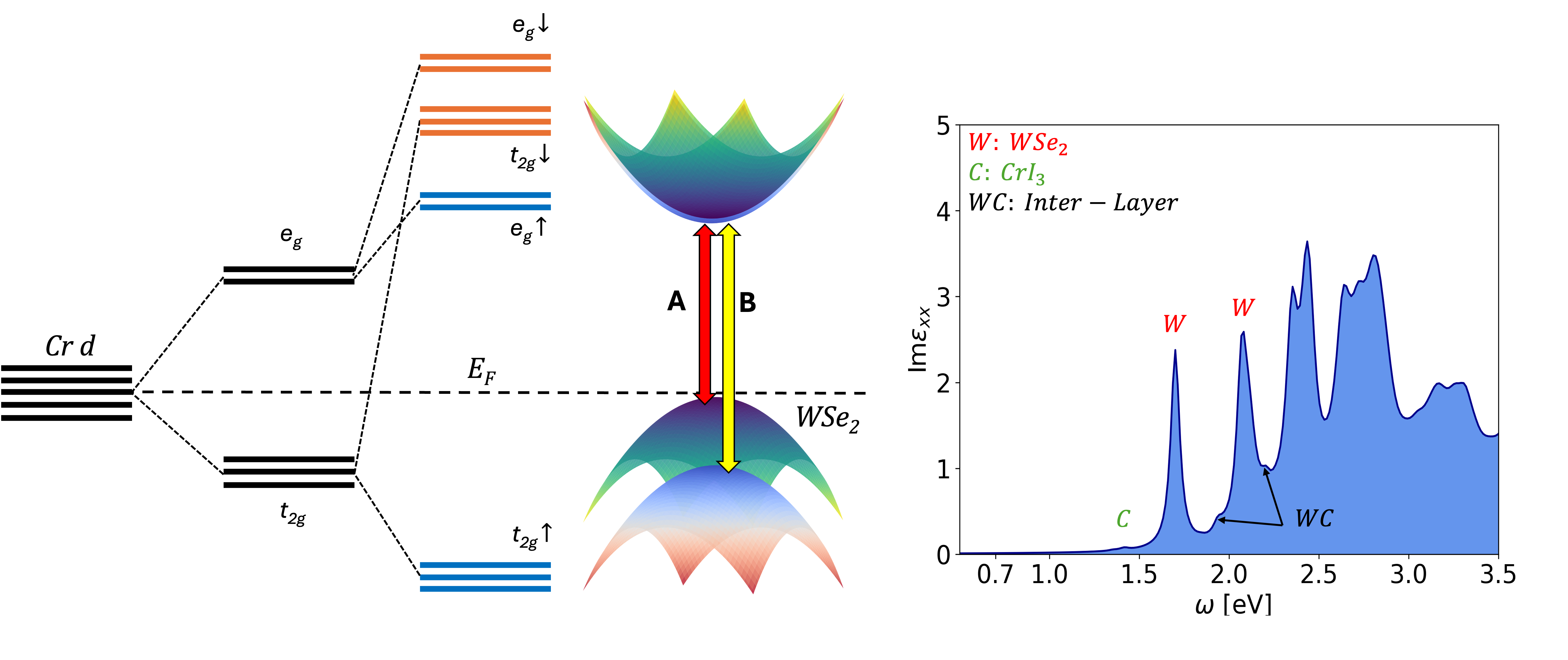}
    \caption{An illustration depicting the origin of different excitonic features in a H-WSe$_2$/CrI$_3$ heterostructure. Cr \textit{d} states energetically split into a degenerate set of three $t_{2g}$ states and two $e_g$ states in the presence of an octahedral crystal field. The relative alignment of CrI$_3$ atomic-like states and the WSe$_2$ bloch states as predicted by $\mathrm{QS}G\hat{W}$ ensures that the primary excitonic features from WSe$_2$ are preserved in both spin channels. At the same time, the important excitonic features from the CrI$_3$ layer are also preserved in the heterostructure. In addition to the excitons originating from the respective layers, a new set of magnetic interlayer %IZ 
    excitons are observed with enhanced oscillator strengths compared to the relatively dark CrI$_3$ excitons and the energies of these excitons are highly robust in both the bilayer and trilayer configurations.}
    \label{fig:summary}
\end{figure}

Here we employ this approach to explore the WSe$_2$/CrI$_3$ interface, in particular what excitons emerge and the impact of proximity-generated %IZ - added
magnetic field on the WSe$_2$ excitons. For this system, the low energy part of the valence and conduction states of the CrI$_3$ is primarily determined by $e_{g}$ and majority-spin $t_{2g}$ states~\cite{wu2019physical,acharya2022real,PhysRevMaterials.6.014008}.  Thus excitonic spin-flip processes are unimportant:  $\mathrm{QS}G\hat{W}$ is sufficient, and augmentation with DMFT 
is not needed. $\mathrm{QS}G\hat{W}$  has been highly successful in describing both the 1-particle~\cite{acharya2021electronic,bianchi2023paramagnetic,watson2024giant} and excitonic states~\cite{acharya2022real,shao2024exciton,datta2024magnon,ruta2023hyperbolic} involving Cr$^{3+}$ in a range of %IZ 2D defined above two-dimensional (
2D %) 
vdW magnets. This gives us confidence that modeling the bilayer and trilayer heterostructures in this work is high enough fidelity to understand the spin-splitting, oscillator strengths and wavefunctions of the excitons generated at the interface owing to proximity effects, %IZ s added
with the caveat that uncertainty in the nuclear configurations leaves some ambiguity we cannot resolve. In Figure~\ref{fig:summary} we summarize the essential finding of our work, where we show the relative alignments between the atomic-like and bloch-like states from CrI$_3$ and WSe$_2$  respectively, and the emergence of the novel interlayer %IZ
excitonic states in addition to the intralayer %IZ
excitons.

%discuss the real issues like small changes in band alignments can dramatically modify the absolute splittings. This band alignment can change due to phonons, strain, interface height. Also at theoretical level additional higher order diagrams can bring in 100-200 meV corrections that can change band alignments leading to undecrtainties in the estimation of the splittings. However, what is robust is the fact we can generate wavfunctions that are of three distict kind and with 3 very different degree of localization (and bohr radius). These states should have different degree of oscillator strengths, coherence and tunability. 
%\begin{figure}[h]
%    \centering
    %\includegraphics[width=0.45\linewidth]{Figures/bilayer_stacking.png}
        %\includegraphics[width=0.45\linewidth]{Figures/trilayer_stacking.png}
%    \includegraphics[width=\textwidth]{Figures/bi_and_tri_stacking.png}
%    \caption{H-WSe$_2$/CrI$_3$ bilayer stacking configuration (C1) of the vdW heterostructure. The unit cell of the heterostructure is marked with black dotted lines. The interlayer distance of 3.53 \AA\ is specified by the distance between the adjacent Se and I planes.}
%    \label{fig:bilayerstacking}
%\end{figure}

\section{Results}
\subsection*{Crystal structure}

The structures were optimized for the bilayer WSe$_2$/CrI$_3$ and the trilayer CrI$_3$($\uparrow$)/WSe$_2$/
CrI$_3$($\uparrow$)[FM], CrI$_3$($\uparrow$)/WSe$_2$/
CrI$_3$($\downarrow$)[AFM] configurations using Quantum Espresso~\cite{giannozzi2009quantum} at density functional theory (DFT) level of theory. The Perdew-Burke-Ernzerhof (PBE)~\cite{perdew1996generalized} functional was used along with ultra-soft pseudo potentials and the atomic positions are relaxed within a fixed cell volume, using a 8$\times$8$\times$1 k-mesh. The kinetic energy cutoff for the wavefunction was set at 100 Ry for all the three structures. Van der Waals coupling between the layers was accounted for by employing Grimme's semi-empirical DFT-D3~\cite{grimme2010consistent} approach.

\begin{figure}[htb!]
    \centering
    \includegraphics[width=\textwidth]{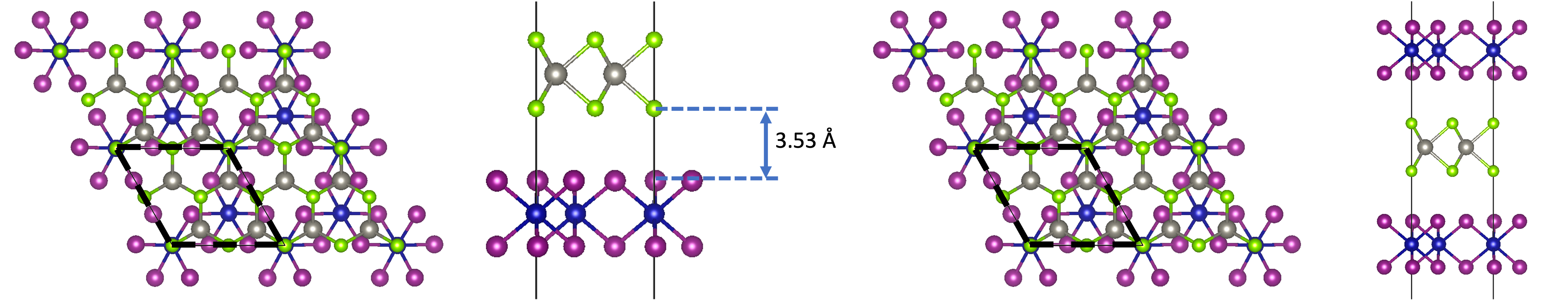}
    \caption{H-WSe$_2$/CrI$_3$ bilayer stacking configuration (C1) of the vdW heterostructure. The unit cell of the heterostructure is marked with black dotted lines. The interlayer distance of 3.53 \AA\ is specified by the distance between the adjacent Se and I planes.}
    \label{fig:bilayerstacking}
\end{figure}

To construct a WSe$_2$/CrI$_3$ heterostructure, a coincident site lattice must be chosen.  Using the fact that the unit cell of CrI$_3$, has a lattice constant roughly twice that of WSe$_2$, we stack a 2$\times$2 WSe$_2$ unit cell over the CrI$_3$.  As the two lattices do not match exactly, some strain is induced. The experimental lattice constant of CrI$_3$ is estimated to be 6.867\AA~\cite{mcguire2015coupling}. Compared to the experimental estimate, the strain in the heterostructure for the CrI$_3$ layer is 3.3\%. \textcolor{black}{Concerning the relative translation of the two sublattices, previous studies~\cite{zhang2019valley} on WSe$_2$/CrI$_3$ bilayer have typically used one of three different symmetric stacking configurations, C1, C2 and C3 as shown in Supplementary Figure 1.} In the C1 configuration, one Cr site lies below the center of the WSe$_{2}$ hexagon and the other below a Se site. C2 is similar except that the second Cr lies beneath the W site.  In C3, the two Cr reside directly underneath W and Se respectively. In this study, we have focused on the C1 configuration as seen in Fig \ref{fig:bilayerstacking}. Relaxing the structure, we find the interlayer distance to be 3.53 \AA. The trilayer heterostructure is constructed by  using a mirror image of the bottom layer for the top layer,
resulting in a sandwiched configuration as seen in Fig \ref{fig:bilayerstacking}. Recent DFT calculations of C1, C2 and C3, predict similar binding energies and therefore any one of them can be observed experimentally~\cite{zhang2019valley,zollner2023strong}.

%\begin{figure}[hbt!] 
%    \includegraphics[scale=0.25]{Figures/tri-afm-lda.pdf}
%    \includegraphics[scale=0.3]{Figures/val-tri-afm-lda.png}
%    \includegraphics[scale=0.3]{Figures/con-tri-afm-lda.png}  \\  
%    \includegraphics[scale=0.25]{Figures/tri-afm-rpa.pdf}
%    \includegraphics[scale=0.3]{Figures/val-tri-afm-rpa.png}
%    \includegraphics[scale=0.3]{Figures/con-tri-afm-rpa.png}\\
%    \includegraphics[scale=0.25]{Figures/tri-afm-bse.pdf} 
%    \includegraphics[scale=0.3]{Figures/val-tri-afm-bse.png}
%    \includegraphics[scale=0.3]{Figures/con-tri-afm-bse.png}
%    \includegraphics[scale=0.6]{Figures/AFM_tri_band_and_dos.png}
%    \caption{(From top to bottom) $\mathrm{QS}G\hat{W}$, QS$GW$, and LDA band structures %IZ for the CrI$_3$($\uparrow$)/WSe$_2$/CrI$_3$($\downarrow$) trilayer structure. The colors in the band plots correspond to W(green), Se(black), Cr(blue) and I(red) respectively. The corresponding partial density-of-states projected onto Cr-$d$, W-$d$, Se-$p$ and I-$p$ states, for the valence and conduction manifolds are shown respectively.}
%    \label{fig:afmtriband}
%\end{figure}
\subsection*{Electronic structure} %IZ Strucutre, if 2nd words not capitalized elsewhere
%\subsubsection*{WSe$_2$/CrI$_3$ bilayer and CrI$_3$/WSe$_2$/CrI$_3$ trilayer}
Energy band structures for the WSe$_2$/CrI$_3$ bilayer at the LDA, QS$GW$ and $\mathrm{QS}G\hat{W}$ levels of theory are summarized in Supplementary Figure 2, while %IZ
Fig.~\ref{fig:afmtriband} and Supplementary Figure 3
%of the main text 
presents the %IZ
corresponding results for the trilayer variants.  As the most important details of the band structure %IZ space 
are similar between the bilayer and trilayer variants, we will use the trilayer results in the main text for representation purposes.
(The %while 
results for all layered and magnetic variants are produced in the Supplementary Information.)
%IZ materials.)
The corresponding band gaps are tabulated in Table~\ref{tab:bandgap}.   At the %IZ
LDA level, a type II band alignment is predicted for the WSe$_2$/CrI$_3$ bilayer, with a band gap of 0.77 eV. The valence band maximum (VBM) is situated on the WSe$_2$ layer whereas the conduction band minimum (CBM) sits on the CrI$_3$ layer. Howeverer, %IZ But 
as is typical of LDA, the band gaps are underestimated for semiconductors~\cite{questaal_paper} and magnets~\cite{acharyaTheoryColorsStrongly2023} and in particular, the band gap for 2D magnets such as Chromium tri-halides is quite severely underestimated~\cite{acharya2021electronic,wu2019physical,molina2020magneto}, Crucially, the orbital character of states at both the valence and the conduction band edge differ from the MBPT results (Fig.~\ref{fig:afmtriband} and Supplementary Figures 2 and 3).
%\begin{table}[h]
%    \centering
%    \begin{tabular}{|c|c|c|c|}
%    \hline
%    System  &  LDA\,(eV) & QS$GW$\,(eV)  & $\mathrm{QS}G\hat{W}$\,(eV)\\
%    \hline
%    WSe$_2$/CrI$_3$  & 0.77 &  2.33   & 2.21 \\
%    CrI$_3$($\uparrow$)/WSe$_2$/CrI$_3$($\uparrow$)  & 0.82 &  2.29   & 2.20 \\
%    CrI$_3$($\uparrow$)/WSe$_2$/CrI$_3$($\downarrow$)  &  0.80 &  2.29   & 2.20  \\
%    \hline
%    \end{tabular}
%    \caption{Band gaps for the WSe$_2$/CrI$_3$, CrI$_3$($\uparrow$)/WSe$_2$/CrI$_3$($\uparrow$) and CrI$_3$($\uparrow$)/WSe$_2$/CrI$_3$($\downarrow$) vdW heterostructures reported in this study.}
%    \label{tab:bandgap}
%\end{table}
 At the QS$GW$ level, states of all relevant orbital character (Cr-\textit{d}, W-\textit{d}, Se-\textit{p}, and I-\textit{p}) move higher in energy relative to the LDA, but the degree of the shift varies with orbital character.  The band gap increases while the band alignment changes from type II into type I. This change in the band alignment can be attributed to the fact that the renormalization of the band structure for the 2D magnet (in this case CrI$_3$) is much larger than in TMDs %IZ TMDCs 
 like WSe$_2$. This is natural since the Cr $t_{2g}$ and $e_{g}$ states are rather non-dispersive and the self-energy corrections for them are greater than the dispersive W-\textit{d} orbitals.

The gap opens up to 2.33~eV with conduction bands of Cr-\textit{d} character moving above bands of W-\textit{d} character, so that the conduction band minimum (CBM) consists mostly of the latter. States of I-\textit{p} character are swept to energy similar to that of the Cr-\textit{d}, and the two hybridize.  With I-\textit{p} now well above W-\textit{d}, the CBM shifts to the K point. The valence region is also significantly altered relative to the LDA, the I-\textit{p} and Se-\textit{p} states that appear prominently about 0.2~eV below the valence band maximum (VBM) in the LDA, are pushed deeper.  The VBM consists of W-\textit{d} character, so at the QS\textit{GW} level both band edges are of W-\textit{d} character, and both occur at the K point.  Thus the QS\textit{GW} and LDA bands are quantitatively and qualitatively different (compare middle and bottom panels Fig.~\ref{fig:afmtriband}). \textcolor{black}{Supplementary Figure 4 illustrates the band gap convergence at the QS$GW$ level w.r.t k-mesh. The QSGW bandstructures for the C2 and C3 stackings closely resemble that of the C1 stacking with band gaps of 2.38 eV and 2.36 eV respectively, as summarized in Supplementary Figure 5.}
%\begin{table}[h]
%    \centering
%    \begin{tabular}{|c|c|c|}
%    \hline
%      State & $|\Delta_{ex}|$/meV (GHz) & Layer contributions \\
%      \hline
%      v$_{K,0}$   & 1.4 (339) & 99.6 WSe$_2$, 0.4 CrI$_3$\\
%      c$_{K,0}$   & 1.4 (339) & 99.7 WSe$_2$, 0.3 CrI$_3$ \\
%      c$_{K,1}$   & - & 73 WSe$_2$, 27 CrI$_3$\\
%      \hline
%    \end{tabular}
%    \caption{Exchange splitting $|\Delta_{ex}|$ at the WSe$_{2}$ layer generated by the proximity field from the CrI$_{3}$ layers, shown for  %IZ $\Delta_{ex}$ of 
%    the top 
    %IZ most 
%    valence band and the lowest %IZ bottom most conduction bands at the K point of the trilayer CrI$_3$/WSe$_2$/CrI$_3$ (FM) configuration. }
    %IZ are shown.}
%    \label{tab:statesplit}
%\end{table}

 QS$GW$ is known to systematically overestimate the band gaps in semiconductors by 10-20\%~\cite{qsgw}. RPA misses the electron-hole binding and hence produces an overly large $W$ that makes the band gaps too large. The missing electron-hole attraction can be diagrammatically included by the addition of excitonic ladder diagrams in the polarizability~\cite{Cunningham2023}.  Carrying the improved polarizability through the
self-consistency cycle makes it possible to describe both the optical properties and the band gap accurately. In the WSe$_2$/CrI$_3$ bilayer system, addition of ladder diagrams in \textit{W} ($\mathrm{QS}G\hat{W}$) reduces the band gap slightly to 2.21~eV. The most important effect of ladders is to reduce splittings between occupied and unoccupied states of Cr-\textit{d} and I-\textit{p} character, but the states close to the band edges are essentially similar to QS\textit{GW} (compare top and middle panels of Fig.~\ref{fig:afmtriband}).

Since the band alignment is of type I and both band edges reside on WSe$_2$ (each with W-\textit{d} character), this leads to a unique platform where A and B excitons of pristine WSe$_2$ are preserved in the heterostructure. The relative alignment between the Cr-\textit{d} and  W-\textit{d} states (from both the valence and conduction sides) becomes the primary mechanism that determines both the magnetic splittings and the formation of a new delocalized, interlayer exciton, to be described below. That absolute magnitudes of the band gaps between CrI$_3$ ($\sim$2.6 eV) and WSe$_2$ ($\sim$2.3 eV) predicted by the theory are similar, has a significant consequence for the excitons that emerge.  Only partial experimental confirmation of this finding is available, however.  From various measurements and theoretical estimates the WSe$_2$ band gap is established to be $\sim$ 2.3~eV~\cite{he2014tightly,hanbicki2015measurement}, but the situation is more ambiguous for CrI$_3$. Reported band gaps vary between 1~eV and 2.6~eV~\cite{kundu2020valence}. This is not a one-off situation. For most 2D magnets, charging effects lead to a large uncertainty in band gap determination not seen in the TMDs. %IZ TMD's. 
In recent works, we showed in collaborations with two ARPES groups~\cite{bianchi2023paramagnetic,watson2024giant}, that the charging effects in another Cr-based 2D magnet Cr$_{2}$Br$_{2}$S$_{2}$ led to 0.5 eV difference in experimental estimation of band gaps from the two groups. $\mathrm{QS}G\hat{W}$'s prediction of 2 eV gap in Cr$_{2}$Br$_{2}$S$_{2}$ was in near perfect agreement with the experiment with least charging effect and the sample closest to pristine conditions~\cite{watson2024giant}. So, while $\mathrm{QS}G\hat{W}$ has an excellent track record in predicting band gaps over a wide range of pristine materials where the gaps are reliably known~\cite{Cunningham2023}, differences in experimental conditions may realize different band gaps owing to their deviations from the pristine conditions. This is particularly important in these WSe$_2$/CrI$_3$ heterostructures since the band gaps of the constituents are very similar and small deviations from the pristine conditions, or assumed structure, or inaccuracies in the theory can lead to shift from type %IZ
I to type %IZ
II alignment. Note that instead of CrI$_3$, if CrBr$_{3}$ was used, this would not be the case since the bonding is more ionic in CrBr$_{3}$ and also the Br valence states are much deeper than the iodine states, which leads to a much larger ($\sim$3.8~eV) band gap~\cite{acharya2021electronic}.

%\begin{figure}[hbt!]
%    \centering
    %\includegraphics[width=0.5\linewidth]{Figures/ImEall.pdf}\includegraphics[width=0.5\linewidth]{Figures/ImE-bil.pdf}
%    \includegraphics[width=\textwidth]{Figures/e2plot.png}
%    \caption{%(IZ left) 
%    Left: Imaginary 
%    part of the dielectric function (Im$\epsilon_{xx}$ for the bilayer and trilayer configurations). %IZ added )
%    The optical response is shown for the perturbing electric field applied along the (100) direction. %IZ (right)
%    Right: Dielectric response of the WSe$_2$/CrI$_3$ bilayer along the (100) direction.}
%    \label{fig:bil_diefn}
%\end{figure}

%\subsubsection*{CrI$_3$/WSe$_2$/CrI$_3$ trilayer: AFM and FM}
The band structures for the AFM trilayer configuration are summarized in Fig.~\ref{fig:afmtriband}. (The corresponding figure for the FM trilayer is presented in Supplementary Figure 3.)  As seen in the bilayer case, LDA predicts a type %IZ type above Type was used, choose one or the other, not both 
%IZ am adding type
II band alignment for both FM and AFM configurations, while QS$GW$ and $\mathrm{QS}G\hat{W}$ predict type I band alignment. 
In the conduction manifold for both AFM and FM, the CBM is dominated by W-\textit{d} states and the Cr-\textit{d} and I-\textit{p} states are swept still further away from the band edges compared to the bilayer as seen in Fig. \ref{fig:afmtriband} and Supplementary Figure 3 respectively. However, the band gaps for the bilayer and trilayer configurations are similar (Table \ref{tab:bandgap}) since the band edges are determined primarily by WSe$_2$. The distinct spin configurations in the tri-layer variants, (FM, AFM), don't change any important details for the crucial band alignments on the valence side, however, the Cr e$_{g}$ state on the conduction side, is slightly closer to the W-\textit{d} van Hove in the vicinity of the CBM for the FM trilayer configuration. This reflects in the computed exchange splitting parameter $\Delta_{ex}$. The proximity field generated at the WSe$_2$ layer by the CrI$_{3}$ magnetic layers produces $\sim$1 meV splitting ($\sim$10 tesla) %IZ magnetic field) 
at the valence and conduction edges at the high symmetry $K$ point. As we move away from the lowest conduction band, the pristine WSe$_2$ conduction states are not present but a new set of states with admixture of both WSe$_2$ and CrI$_3$ characters are observed, which makes the analysis of exchange splitting non-trivial for such states. As reported in Table \ref{tab:statesplit}, the second conduction state at the K point is observed to have a mixture of roughly 73\% of WSe$_2$ and 27\% of CrI$_3$.   %However, the overlap of the Cr e$_{g}$ conduction state with the second 
%lowest
%IZ most 
%conduction band from WSe$_{2}$ imply that the $\Delta_{ex}$ is significantly enhanced for that particular state. As we show later, such  differences can have a dramatic impact on the exciton splitting, particularly promising is the the possibility to realize a large magnetic splitting of the B exciton state in the trilayer FM configuration. % (no DFT values available for comaprison for FM trilayer) We note that the $\Delta_{ex}$ computed from our approach for the band edge states is significantly smaller than the predictions from DFT~\cite{}. CITATION MISSING %IZ
%IZ density functional theory~\cite{}.  DFT was already used many times

%\begin{figure}[hbt!]
%    \centering
 %   (a)
 %   \includegraphics[scale=0.5]{Figures/bil-ex-cri3.png} \\
 %   (b)
 %   \includegraphics[scale=0.5]{Figures/bil-ex-wse2.png} \\
 %   (c)
 %   \includegraphics[scale=0.5]{Figures/bil-ex-cw.png} \\
 %   \includegraphics[width=\textwidth]{Figures/types_of_exctions.png}
 %   \caption{Exciton wavefunction analysis for (a) Purely CrI$_3$ exciton (b) purely WSe$_2$ exciton and (c) 
 %   Interlayer %IZ interlayer Note: many times interlayer was used, please choose one of them
 %   (WSe$_2$/CrI$_3$) exciton in WSe$_2$/CrI$_3$ bilayer heterostructure. The images on the left represent the exciton hole wavefunctions in the real space. On the right, the corresponding pie-charts indicate the integrated spectral weights decomposed over the layers. `X' marks the location of the electron/hole while plotting the exciton hole/electron wavefunction.}
%    \label{fig:bil-excitons}
%\end{figure}

\subsection*{Excitonic spectrum}
%\subsubsection*{WSe$_2$/CrI$_3$ bilayer and CrI$_3$/WSe$_2$/CrI$_3$ trilayer}
The nonmagnetic %IZ nonmagnetic seems correct, but many times nonmagnetic was used, please use one of them
Wannier-Mott excitons, A and B, in WSe$_{2}$ have binding energies E$_{b}$ of 0.79~eV and 0.37~eV respectively\cite{hanbicki2015measurement}.  Exciton binding energies calculated within the $\mathrm{QS}G\hat{W}$ framework  are in close agreement with the experimental observations. The A exciton originates from the top-most valence band and bottom-most conduction band states taken from a small region $k$ in the vicinity of the $K$ point, whereas the B exciton originates from the second valence band and bottom most conduction band\cite{zhao2013evolution,he2014tightly}. The van Hove features corresponding to these states can be observed in the partial density of state plots in Fig.~\ref{fig:afmtriband}. Note that the Cr-\textit{d} states on the valence side are far from the (hole-like) van Hove features that correspond to A and B excitons while the Cr-\textit{d} state from the conduction state is close to the electron-like state that takes part in forming both A and B excitons. In both the bilayer and trilayer, the energies of the A and B excitons remain consistent(Fig.~\ref{fig:bil_diefn}), and indeed these excitons are largely  confined to WSe$_{2}$  for the bilayer and FM trilayer (Fig.~\ref{fig:bil-excitons} and Supplemetary Figure 6) and almost exclusively so for the AFM trilayer (Fig.~\ref{fig:Bex_tri} and Supplementary Figure 7). \textcolor{black}{Supplementary Figures 8 and 9 summarizes the convergence of the optical spectrum computed within the $\mathrm{QS}G\hat{W}$ framework against the active space and k-mesh.} In addition to the nonmagnetic WSe$_{2}$ excitons, we also find all the CrX$_{3}$ magnetic excitons found in experimental optical absorption spectra~\cite{RevModPhys.83.705,mcguire2015coupling}. The three important Frenkel-like CrX$_{3}$ excitons~\cite{ghosh2023magnetic,acharya2022real,wu2019physical} which are relevant for our purposes of proximity effect are observed at 1.45 eV, 1.85 eV and 2.1 eV, respectively. The electron- and hole-states that form these excitons, in sharp-contrast with the WSe$_{2}$ excitons, are delocalized over several bands and over the entire Brillouin zone. As a result, in real space, %IZ - hyphen removed
the CrI$_{3}$ excitons extend over only a few Angstroms to a maximum of a nanometer while the WSe$_{2}$ excitons extend over several 10's of nanometers. This has a remarkable consequence in the distinct possibilities for the exciton formation in the heterostructures. Note that the lack of screening and the partially filled character of the Cr-\textit{d} states make them highly non-dispersive which makes the CrI$_{3}$ $dd$ excitons significantly localized in real space. The additional weak screening that WSe$_{2}$ brings to the heterostructure can not affect the energy and localization of these Frenkel excitons. However, the presence of weak interlayer %IZ hyphen removed
dipole means that the stringent optical selection rules for the onsite $dd$ transitions are partly relaxed and the oscillator strengths of these Frenkel excitons increase. This principle can be exploited in general to better identify and read out information from these relatively dark $dd$ excitons in a TMD-magnet %IZ Magnet 
heterostructure framework. 

%\begin{table}[hbt!]
%    \centering
%    \begin{tabular}{|c|c|c|}
%    \hline
%      Structure   & A exciton/ meV (GHz) & B exciton/ meV (GHz) \\
%      \hline
%      WSe$_2$/CrI$_3$ & 0.42(101) & 1.1 (266) \\
%      CrI$_3$/WSe$_2$/CrI$_3$ (FM) & 0.20 (48) & 5.6 (1354)\\
%      CrI$_3$/WSe$_2$/CrI$_3$ (AFM) & 0.13 (31) & 0.52 (126) \\
%      \hline
%    \end{tabular}
%    \caption{Spin-splitting of the A and B excitons of WSe$_2$ in the different heterostructure configurations at $\mathrm{QS}G\hat{W}$ level of theory.}
%    \label{tab:splitting}
%\end{table}

In addition to these purely nonmagnetic Wannier-Mott and magnetic Frenkel excitons which belong to their respective layers, we also observe series of entirely new excitons which have interlayer %IZ
character. %IZ Remarkably, 
% we have "the most remarkable" just below
Surprisingly, %IZ 
the energy sequence of these three important categories of excitons generated in our heterostructures are robust across all bilayer, trilayer and their magnetic variants. The most remarkable features of these interlayer %IZ
excitons are their systematic appearance in between the A and the B excitons and their enhanced brightness. In this case, the interlayer %IZ
dipole has a greater impact on brightness, because the exciton is dispersed over all the layers and senses the dipole more strongly.  The calculated oscillator strength  is nearly twenty times that of the intralayer %IZ
Frenkel excitons in CrI$_{3}$. Note that this interlayer %IZ
dipole is the secondary mechanism to the primary one from the intralayer %IZ
Iodine ligands ($pd$ intralayer %IZ
hybridization between the Cr-\textit{d} and I-\textit{p} states) that gives oscillator strength to these otherwise {\it dark} excitons~\cite{acharya2022real}.  Typically, without any coupling with the lattice, the relative oscillator strength of the A exciton and 1.45 eV CrI$_{3}$ exciton is $\sim 500$. The interlayer %IZ
exciton, hence, is significantly darker ($\sim$ two orders of magnitude) compared to the A exciton. \textcolor{black}{The brightness of the interlayer exciton observed in our case is comparable to the ones observed in TMD heterobilayers~\cite{yu2015anomalous,baranowski2017probing,rivera2018interlayer,wu2018theory}. We also observe the A2s exciton around 1.83 eV as was reported before~\cite{he2014tightly}, which is darker than the primary A exciton by roughly two orders of magnitude and is slightly darker than our observed interlayer exciton at 1.84 eV. While the dipole strength of the A2s exciton is larger in plane compared to the interlayer exciton, the interlayer exciton is brighter along the out-of-plane direction. These two excitons of very different origin being within only 10 meV of each other does make it difficult for experiments to resolve them. Having said that their primary dipoles are along different polarization directions which might facilitate their detection.} This principle can be exploited in general to identify, control and read out information from these interlayer %IZ
excitons (a principle that is absent from the `dark' $dd$ excitons) in a TMD-magnet %IZ Magnet 
heterostructure framework. We believe, its robustness across all layered variants (binding energy $\sim$400~ meV), significant brightness and magnetic character make these excitons attractive for multiple photonic and spintronic applications. \textcolor{black}{To be able to predict the radiative and non-radiative components of the lifetimes of these excitons would be an important step in advancing this field. However, the non-radiative lifetime effects can only be incorporated in our theoretical approach once the exciton-phonon coupling is included. Further, to be able to compute the radiative lifetimes one needs a dynamical self energy and/or a dynamical vertex W. The quasiparticlization process and the solution of the Bethe-Salpeter Equation (BSE) in the static approximation for W mean both the mechanisms are absent from our present approach.} 
%\begin{figure}[hbt!]
%    \centering
    %\includegraphics[scale=0.4]{Figures/exB_tri.png}
%    \includegraphics[scale=0.25]{Figures/exBwse2-tri-fm.png}\includegraphics[scale=0.25]{Figures/exBwse2-tri-afm.png}
%    \includegraphics[width=\textwidth]{Figures/exB_wse2_afm_and_fm.jpg}
%    \caption{Projection of exciton spectral weights onto onsite and intersite dipoles with dipole contributions summed over layers for B exciton of WSe$_2$ in CrI$_3$($\uparrow$)/WSe$_2$/CrI$_3$($\uparrow$) (left) and CrI$_3$($\uparrow$)/WSe$_2$/CrI$_3$($\downarrow$) (right) respectively. The degree of interlayer %IZ
%    component is observed to be directly proportional to the extent of the spin-splitting.}
%    \label{fig:Bex_tri}
%\end{figure}

Further, we observe moderate magnetic splittings (Table \ref{tab:splitting}) of the A and B excitons, the absolute magnitudes of which are sensitive to the details of van Hove alignments between the Cr-\textit{d} and W-\textit{d} states. 
The largest magnetic splitting generated by the proximity field is observed in the trilayer FM configuration while the splittings in the bilayer and trilayer AFM configurations are very similar and are rather small. Overall, the magnetic splittings are between a few 10's of GHz to 100 GHz in bilayer and trilayer AFM configurations while it is about 1000 GHz in the trilayer FM. The remarkable enhancement in the magnetic splitting of the B exciton has its roots in the fact that the electron-like Cr-\textit{d} van Hove is closer to the WSe$_{2}$ electron-like van Hove that constitutes the B exciton in FM compared to AFM. The absolute shift in Cr-d electron-like van Hove state is only about 20 meV going from the AFM to FM but that is enough to create a large splitting of the B exciton. Note that although the splitting is about 1000 GHz, these magnetically split peaks may still be within the full width at half maximum of the B exciton making them difficult to distinguish from the original nonmagnetic peak. We explore further the excitonic wavefunctions of the B peak in AFM and FM configurations which reveal that the WSe$_{2}$-CrI$_{3}$ interlayer %IZ
dipole element is finite in FM (Figs. ~\ref{fig:Bex_tri} and Supplementary Figure 6) while it is nearly zero in AFM, in consistency with our density of states analysis. \textcolor{black}{In case of the A exciton, the spin splitting is observed to reduce slightly as we go from the bilayer to the FM trilayer configuration. Despite the fact that the absolute value of splitting can be sensitive to the relative alignment of Cr $e_g$ states and the WSe$_2$ conduction states, we note that the small values associated with splittings of the A exciton makes it complicated for the theory to resolve it reasonably well.} Further, for the AFM trilayer configuration, a finite spin-splitting of both the A and B excitons is observed despite net zero magnetization arising from the interlayer anti-ferromagnetic arrangement of the Cr-\textit{d} spins. This observation aligns with the recent findings of Hu et. al~\cite{hu2020manipulation} and Zhang et al.~\cite{zhang2019valley}. However, the general consensus that emerges from these calculations is that the magnetic splitting is 
%IZ  I am trying to make this statement a little
%softer, please remove it, if you dislike it
%IZ extremely 
very sensitive to the band alignments between the TMD and the magnet and small details can significantly %IZ
change its magnitude 
%IZ the numbers drastically 
posing a challenge %IZ great difficulty 
for in-situ control of the splittings in 
these heterostructures. %IZ framework.  

\section{Discussion}
%IZ In summary, we observe
Our findings reveal that the considered %IZ
vdW TMD-magnet heterostructures are excellent platforms for in-situ generation and control of novel interlayer %IZ
excitonic states. These excitons have the best of both worlds; the bright and dipolar character of TMD excitons and on-site magnetic character of the vdW magnetic excitons. The remarkably consistent appearance of these excitons between the A and the B excitons and their %IZ its 
significantly enhanced brightness compared to the intralayer %IZ 
magnetic $dd$ excitons make them attractive for photonic applications. These excitonic states are more tunable than a Frenkel exciton and are sufficiently bright that they can be identified and their information can be read out. Since %IZ, 
the entire CrX$_{3}$ series contains a series of magnetic excitons in the visible range and with energies similar to the A and the B excitons, we expect that %IZ believe, 
the principles we establish in the current work are more generally applicable for a large class of heterostructures formed with varied combinations between these magnets and other TMDs. %IZ TMD's. 
However, our work also provides some intuitive guidance %pictures 
for where we can expect deviations. For %IZ
example, if CrBr$_{3}$ was used instead %IZ order
of CrI$_{3}$ in the heterostructure, we would expect more remarkable enhancements in the brightness of these interlayer %IZ
excitons compared to their intralayer %IZ
counterparts. There are primarily two reasons for that (the Cr-Br bonding is more ionic compared to Cr-I bonding and,  hence, CrBr$_{3}$ intralayer %IZ 
excitons are darker than the CrI$_{3}$ excitons); (a) %once the heterostructure is formed, the lowering of symmetry due to interlayer %IZ dipole formations
\textcolor{black}{once the heterostructure is formed, the breaking of time-reversal symmetry due to the presence of magnetic Cr$^{3+}$ ions and slight lattice mismatch would have more dramatic effect on the brightness of the emergent excitonic states, }(b) Br being a lighter element compared to I, it would induce lattice fluctuations which can further lead to enhancement in brightness of the excitons. However, contrary to the naive expectations of larger proximity field generated splittings by more ionic CrBr$_{3}$, we believe that CrBr$_{3}$ would lead to smaller magnetic splittings of A and B exciton states since the heavy electron like states of CrBr$_{3}$ are more than an electron volt away from the electron like states that form A and B excitons. In such cases, for magnetic splittings we would have to rely on the proximity of the hole-like states from the TMD and magnetic layers of the heterostructures. However, since Br states are deeper than Iodine, even on the valence side, we believe, that the CrBr$_{3}$ hole-like states will be well split off from the hole-like states relevant for A and B excitons. The comparative discussion of CrBr$_{3}$ and CrI$_{3}$ is exploratory in nature and provides an intuitive picture for what to expect in different combinations for the heterostructure. However, as we show in our calculations, since magnetic splittings are sensitive to the band alignments between the %IZ
TMD and the %IZ 
magnet, those numbers can fluctuate between a few GHz and several 1000 GHz due to factors like, interlayer %IZ
spacing, strain and phonon coupling at the interface. While this uncertainty brings in opportunities for further research, we believe
that %IZ, 
the robustness of the novel interlayer %IZ
excitons is the most promising outcome of the proximity effect generated in the heterostructure framework.

\section*{methods}
\subsection*{Computational methodology} %The Quasiparticle Self-Consistent GW approximation~\cite{qsgw,questaal_paper} is a self-consistent form of Hedin's GW approximation. In contrast to conventional $GW$ implementations, QS$GW$ modifies the charge density and is determined by a variational principle~\cite{variational}.  A great majority of discrepancies with the experimental band gap %IZ space added can be traced to omission of electron-hole interactions in the RPA polarizability.  By adding ladders to the RPA, electron-hole effects are taken into account.  Generating \textit{W} with ladder diagrams has several consequences; most importantly, perhaps, screening is enhanced and \textit{W} reduced.  This in turn reduces fundamental band gaps %IZ space and also valence bandwidths.  The importance of self-consistency in both QS$GW$ and $\mathrm{QS}G\hat{W}$ for different materials have been explored \cite{acharya2021importance}.  Of course, excitons which form the main topic of this work appear only as a result of these interactions.
\textcolor{black}{Local density approximation (LDA), QS$GW$~\cite{qsgw} and $\mathrm{QS}G\hat{W}$~\cite{Cunningham2023} calculations were carried out in the Questaal~\cite{questaal_paper} electronic structure suite. The static quasi-particle self-energy $\Sigma^0(\mathbf{k})$ is generated on a 12$\times$12$\times$1 k-mesh and the dynamical self-energy ($\Sigma(\mathbf{k})$) is generated on a 8$\times$8$\times$1 k-mesh. In each iteration of the QS$GW$ ($\mathrm{QS}G\hat{W}$) self-consistency cycle, $\Sigma^0(\mathbf{k})$ and the charge density are both updated.  Iterations continue until self-consistency in both is reached. The convergence threshold for the self-energy is set at $2{\cdot}10^{-5}$ Ry for the RPA self-energy and $4{\cdot}10^{-5}$ for the Bethe-Salpeter equation (BSE) %IZ BSE 
self-energy. For the bilayer system, the BSE self-energy is generated using 33 valence and 23 conduction states in the active space respectively. For the trilayer, the BSE self-energy is generated with 36 valence and 26 conduction states in the active space.
\subsection*{Exciton wavefunction analysis} 
The spectral weight ($w$) for a given dipole pair is given by the following equation,
\begin{equation}
    w_{a^{-}b^{+},n} = \int_{r_e}\int_{r_h}B(a,r_e)B(a,r_h)\psi(r_e,r_h,n)dr_edr_h
\end{equation}
where `B' functions are atomic-like localized functions centered on atoms `a' and `b' respectively. For $a=b$, the weight gives rise to onsite spectral weights and $a\neq b$ case corresponds to the inter-site spectral weights. The computed weights are then summed over the electron- and hole- sites belonging to respective layers to compute the final dipole contributions coming from respective layers.}

\section*{Data Availability} 
All the input file structures and the command lines to launch calculations are rigorously explained in the tutorials available on the Questaal webpage under the terms of the AGPLv3 license. All data needed to reproduce the results from the paper are available here on  \href{https://zenodo.org/records/16916326?preview=1&token=eyJhbGciOiJIUzUxMiJ9.eyJpZCI6ImE4M2ZiMTRhLTFjYjAtNGU2ZC04ZDI5LTQ1MTRiODU0YjI2YyIsImRhdGEiOnt9LCJyYW5kb20iOiJmZDEyYTE4OTQ2MTBkYWFiMTZmY2RjZGVhNDAxZTEyOCJ9.4Yta3uiAP-FYw4pDP-mKTphm-3992eluaEPv1VvYGxIhaWEGQXk61pJuEOxOueSEeLMBP5AWW6SNswSTTvYceg}{Zenodo}. Additional data requests should be made to the corresponding authors of the paper. 

\section{Acknowledgments}
This work was authored by the National Renewable Energy Laboratory for the U.S. Department of Energy (DOE) under Contract No. DE-AC36-08GO28308. Funding was provided by the Computational Chemical Sciences program within the Office of Basic Energy Sciences, U.S. Department of Energy. I\v{Z} was supported by the U.S. Department of Energy, Office of Science, Basic
Energy Sciences under Award No. DE-SC0004890. We acknowledge the use of the National Energy Research Scientific Computing Center, under Contract No. DE-AC02-05CH11231 using NERSC award BES-ERCAP0021783 and we also acknowledge that a portion of the research was performed using computational resources sponsored by the Department of Energy's Office of Energy Efficiency and Renewable Energy and located at the National Renewable Energy Laboratory and computational resources provided by the Oakridge leadership Computing Facility. The views expressed in the article do not necessarily represent the views of the DOE or the U.S. Government. The U.S. Government retains and the publisher, by accepting the article for publication, acknowledges that the U.S. Government retains a nonexclusive, paid-up, irrevocable, worldwide license to publish or reproduce the published form of this work, or allow others to do so, for U.S. Government purposes. %

\section*{Author Contributions}
MT performed the calculations. SA and MT analyzed the results and drafted the paper. MvS and DP provided the necessary software support. All authors discussed the results and contributed to the draft. 

\section*{Competing Interests Statement}
The authors declare no competing interests.

%\bibliographystyle{ieeetr}

%\bibliographystyle{ieeetr}
%\bibliography{theory,experiments}

\newpage

% Figure 1

\newpage

% Figure 2

\newpage

% Figure 3

\begin{figure}[htb!] 
    \includegraphics[width=\textwidth]{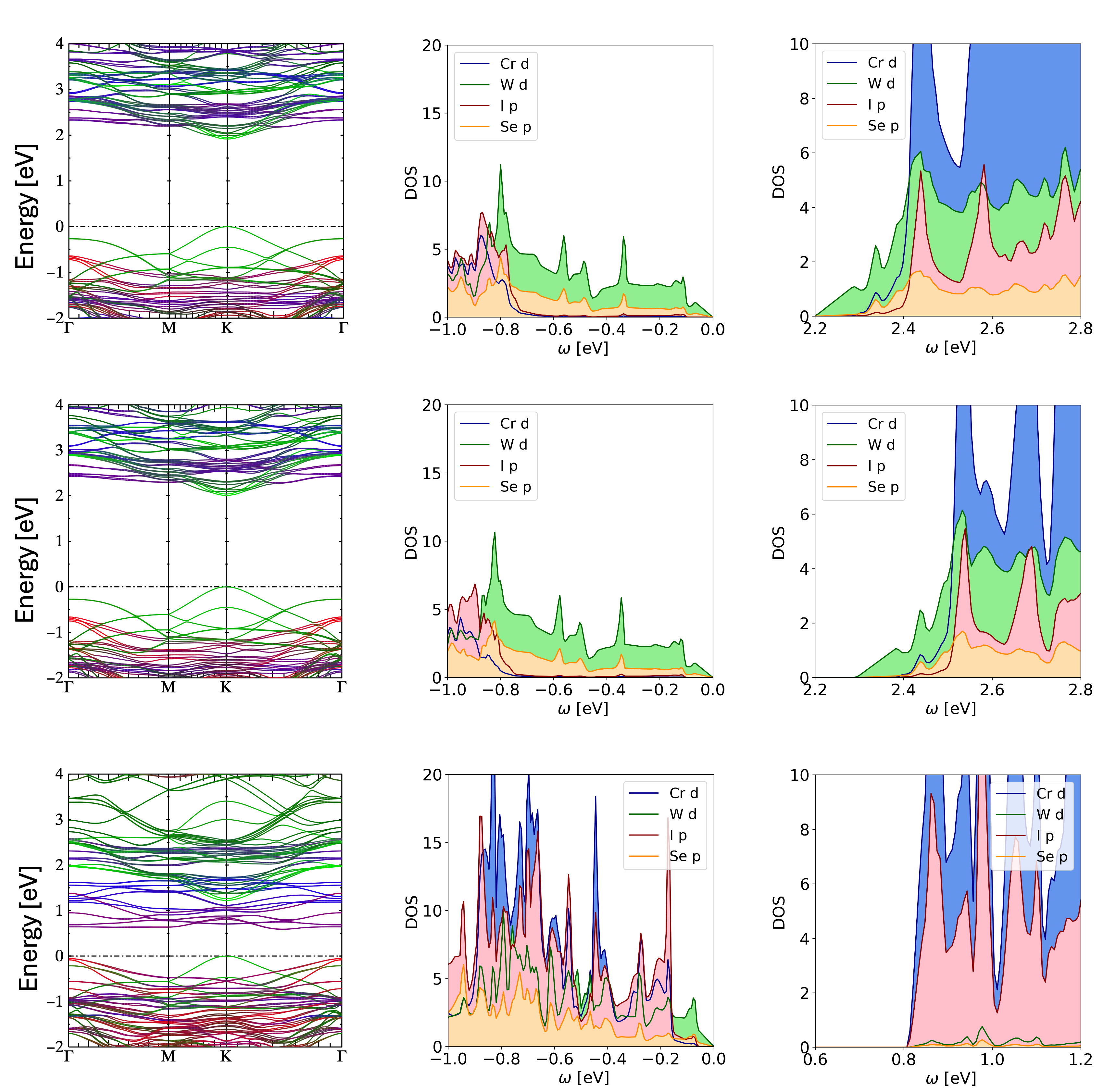}
    \caption{(From top to bottom) $\mathrm{QS}G\hat{W}$, QS$GW$, and LDA band structures %IZ space
    for the CrI$_3$($\uparrow$)/WSe$_2$/CrI$_3$($\downarrow$) trilayer structure. The colors in the band plots correspond to W(green), Se(black), Cr(blue) and I(red) respectively. The corresponding partial density-of-states projected onto Cr-$d$, W-$d$, Se-$p$ and I-$p$ states, for the valence and conduction manifolds are shown respectively.}
    \label{fig:afmtriband}
\end{figure}

\newpage

% Table 1

\begin{table}[htb!]
    \centering
    \begin{tabular}{|c|c|c|c|}
    \hline
    System  &  LDA\,(eV) & QS$GW$\,(eV)  & $\mathrm{QS}G\hat{W}$\,(eV)\\
    \hline
    WSe$_2$/CrI$_3$  & 0.77 &  2.33   & 2.21 \\
    CrI$_3$($\uparrow$)/WSe$_2$/CrI$_3$($\uparrow$)  & 0.82 &  2.29   & 2.20 \\
    CrI$_3$($\uparrow$)/WSe$_2$/CrI$_3$($\downarrow$)  &  0.80 &  2.29   & 2.20  \\
    \hline
    \end{tabular}
    \caption{Band gaps for the WSe$_2$/CrI$_3$, CrI$_3$($\uparrow$)/WSe$_2$/CrI$_3$($\uparrow$) and CrI$_3$($\uparrow$)/WSe$_2$/CrI$_3$($\downarrow$) vdW heterostructures reported in this study.}
    \label{tab:bandgap}
\end{table}

\newpage

% Table 2

\begin{table}[htb!]
    \centering
    \begin{tabular}{|c|c|c|}
    \hline
      State & $|\Delta_{ex}|$/meV (GHz) & Layer contributions \\
      \hline
      v$_{K,0}$   & 1.4 (339) & 99.6 WSe$_2$, 0.4 CrI$_3$\\
      c$_{K,0}$   & 1.4 (339) & 99.7 WSe$_2$, 0.3 CrI$_3$ \\
      c$_{K,1}$   & - & 73 WSe$_2$, 27 CrI$_3$\\
      \hline
    \end{tabular}
    \caption{Exchange splitting $|\Delta_{ex}|$ at the WSe$_{2}$ layer generated by the proximity field from the CrI$_{3}$ layers, shown for  %IZ $\Delta_{ex}$ of 
    the top 
    %IZ most 
    valence band and the lowest %IZ bottom most 
    conduction bands at the K point of the trilayer CrI$_3$/WSe$_2$/CrI$_3$ (FM) configuration. }
    %IZ are shown.}
    \label{tab:statesplit}
\end{table}

\newpage

% Figure 4

\begin{figure}[htb!]
    \centering
    \includegraphics[width=\textwidth]{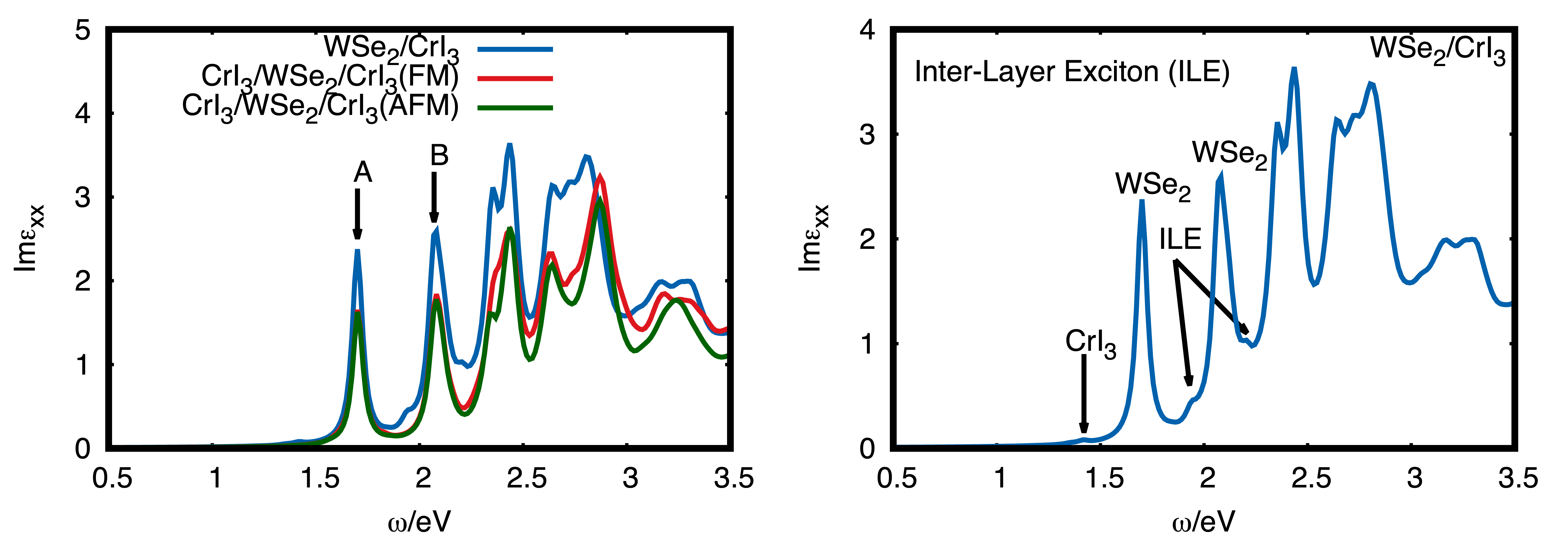}
    \caption{%(IZ left) 
    Left: Imaginary 
    part of the dielectric function (Im$\epsilon_{xx}$ for the bilayer and trilayer configurations). %IZ added )
    The optical response is shown for the perturbing electric field applied along the (100) direction. %IZ (right)
    Right: Dielectric response of the WSe$_2$/CrI$_3$ bilayer along the (100) direction.}
    \label{fig:bil_diefn}
\end{figure}

\newpage

% Figure 5

\begin{figure}[htb!]
    \centering
 %   (a)
 %   \includegraphics[scale=0.5]{Figures/bil-ex-cri3.png} \\
 %   (b)
 %   \includegraphics[scale=0.5]{Figures/bil-ex-wse2.png} \\
 %   (c)
 %   \includegraphics[scale=0.5]{Figures/bil-ex-cw.png} \\
    \includegraphics[width=\textwidth]{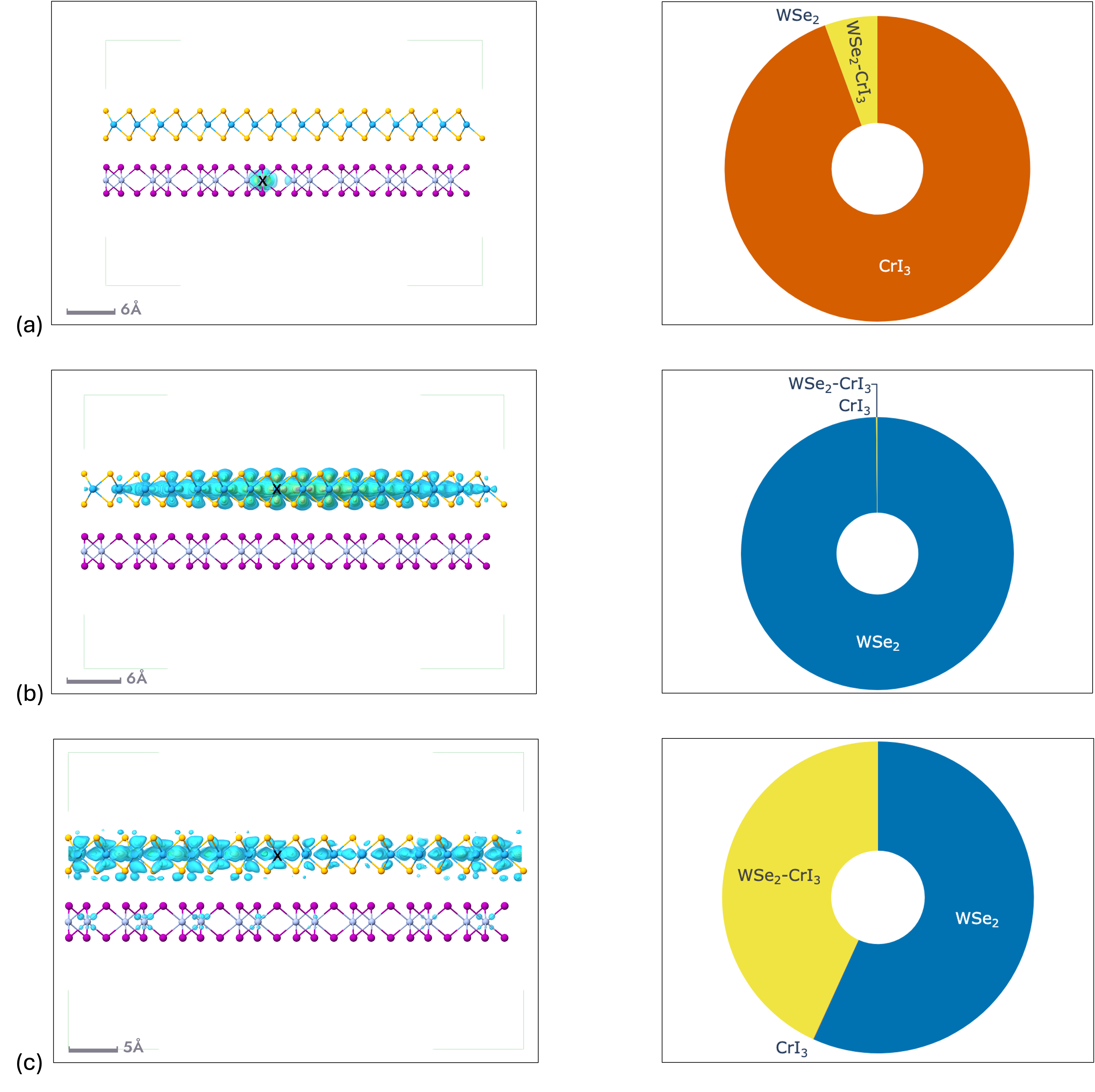}
    \caption{Exciton wavefunction analysis for (a) Purely CrI$_3$ exciton (b) purely WSe$_2$ exciton and (c) 
    Interlayer %IZ interlayer Note: many times interlayer was used, please choose one of them
    (WSe$_2$/CrI$_3$) exciton in WSe$_2$/CrI$_3$ bilayer heterostructure. The images on the left represent the exciton hole wavefunctions in the real space. On the right, the corresponding pie-charts indicate the integrated spectral weights decomposed over the layers. `X' marks the location of the electron/hole while plotting the exciton hole/electron wavefunction.}
    \label{fig:bil-excitons}
\end{figure}

\newpage

% Figure 6

\begin{figure}[htb!]
    \centering
    \includegraphics[width=\textwidth]{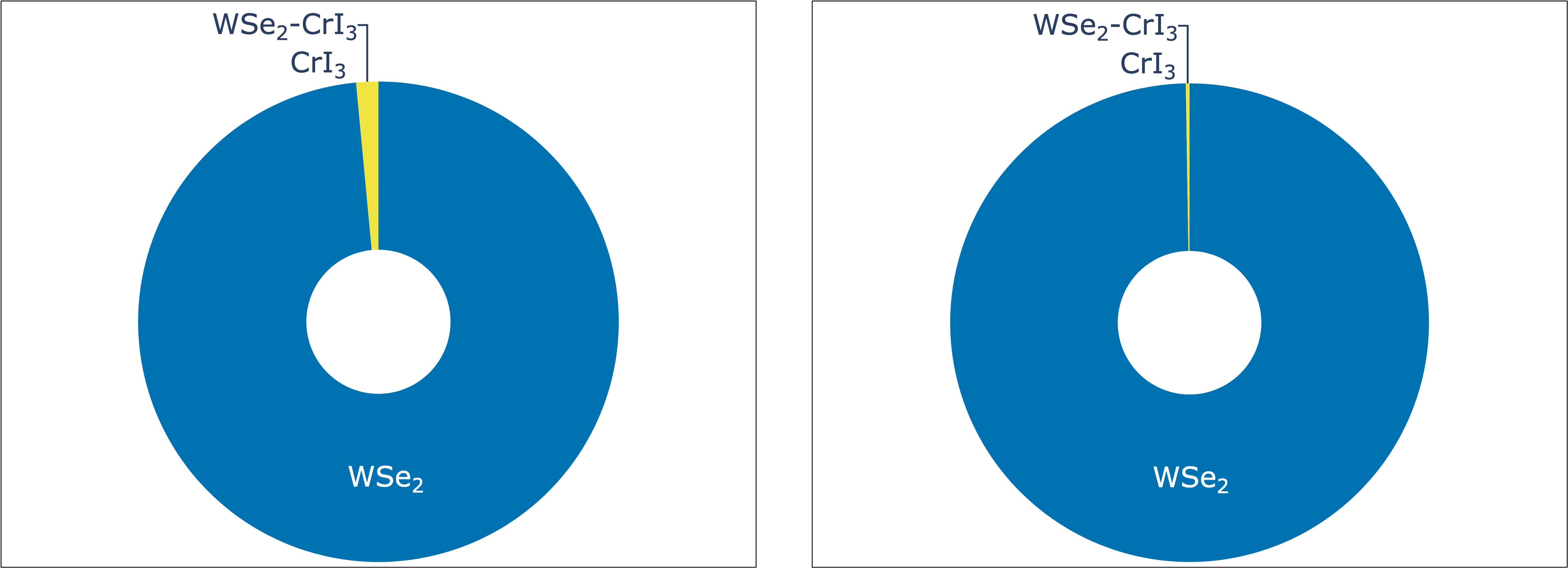}
    \caption{Projection of exciton spectral weights onto onsite and intersite dipoles with dipole contributions summed over layers for B exciton of WSe$_2$ in CrI$_3$($\uparrow$)/WSe$_2$/CrI$_3$($\uparrow$) (left) and CrI$_3$($\uparrow$)/WSe$_2$/CrI$_3$($\downarrow$) (right) respectively. The degree of interlayer %IZ
    component is observed to be directly proportional to the extent of the spin-splitting.}
    \label{fig:Bex_tri}
\end{figure}

\newpage

% Table 3

\begin{table}[hbt!]
    \centering
    \begin{tabular}{|c|c|c|}
    \hline
      Structure   & A exciton/ meV (GHz) & B exciton/ meV (GHz) \\
      \hline
      WSe$_2$/CrI$_3$ & 0.42(101) & 1.1 (266) \\
      CrI$_3$/WSe$_2$/CrI$_3$ (FM) & 0.20 (48) & 5.6 (1354)\\
      CrI$_3$/WSe$_2$/CrI$_3$ (AFM) & 0.13 (31) & 0.52 (126) \\
      \hline
    \end{tabular}
    \caption{Spin-splitting of the A and B excitons of WSe$_2$ in the different heterostructure configurations at $\mathrm{QS}G\hat{W}$ level of theory.}
    \label{tab:splitting}
\end{table}

\end{document}